# Identifying False Content and Hate Speech in Sinhala YouTube Videos by Analyzing the Audio

W. A. K. M. Wickramaarachchi
*Department of Information Technology*
*Faculty of Computing*
*Sri Lanka Institute of Information Technology*
Malabe, Sri Lanka
wickramaarachchikm@gmail.com

Sameeri Sathsara Subasinghe
*Department of Information Technology*
*Faculty of Computing*
*Sri Lanka Institute of Information Technology*
Malabe, Sri Lanka
sameerisubasinghe@gmail.com

K. K. Rashani Tharushika Wijerathna
*Department of Information Technology*
*Faculty of Computing*
*Sri Lanka Institute of Information Technology*
Malabe, Sri Lanka
rashanitharushika99@gmail.com

A. Sahashra Udani Athukorala
*Department of Information Technology*
*Faculty of Computing*
*Sri Lanka Institute of Information Technology*
Malabe, Sri Lanka
sahashraudani@gmail.com

Lakmini Abeywardhana
*Department of Information Technology*
*Faculty of Computing*
*Sri Lanka Institute of Information Technology*
Malabe, Sri Lanka
lakmini.d@sliit.lk

A. Karunasena
*Department of Information Technology*
*Faculty of Computing*
*Sri Lanka Institute of Information Technology*
Malabe, Sri Lanka
anuradha.k@sliit.lk

*Abstract*—YouTube faces a global crisis with the dissemination of false information and hate speech. To counter these issues, YouTube has implemented strict rules against uploading content that includes false information or promotes hate speech. While numerous studies have been conducted to reduce offensive English-language content, there's a significant lack of research on Sinhala content. This study aims to address the aforementioned gap by proposing a solution to minimize the spread of violence and misinformation in Sinhala YouTube videos. The approach involves developing a rating system that assesses whether a video contains false information by comparing the title and description with the audio content and evaluating whether the video includes hate speech. The methodology encompasses several steps, including audio extraction using the Pytube library, audio transcription via the fine-tuned Whisper model, hate speech detection employing the distilroberta-base model and a text classification LSTM model, and text summarization through the fine-tuned BART-Large-XSUM model. Notably, the Whisper model achieved a 48.99\% word error rate, while the distilroberta-base model demonstrated an F1 score of 0.856 and a recall value of 0.861 in comparison to the LSTM model, which exhibited signs of overfitting.

*Keywords*— Hate speech detection, Automatic speech recognition, False content detection

I. INTRODUCTION

Established in 2005, YouTube is a global platform for online video sharing. Since then, it has become one of the world's biggest and most well-known websites, with more than 2.6 billion users actively using it monthly. The ability to publish, distribute, and watch videos for free is one of YouTube's most valuable aspects. As a result, it has become an effective instrument for communication, education, and entertainment.

There is a vast range of content on YouTube, including vlogs, videos in various music genres, news, films, documentaries, live streams, and more with which the viewers can interact. Also, viewers can subscribe to their favorite channels, which will alert them when new videos are uploaded. Furthermore, viewers can like, share, and comment on these videos. YouTube has evolved into more than just a platform for content producers; it has also enabled numerous individuals to succeed in making a living off of their channels through sponsorships, advertising income, and sales of items.

YouTube has generated some controversy, with concerns about copyright infringement, false information, and hazardous content. The spread of false content and hate speech has become a significant threat worldwide; uploaders have found YouTube to share incorrect information using clickbait content and spread violence using hate speech, significantly impacting the video's trustworthiness negatively. This is because Clickbait content, designed to attract attention and generate clicks, often contains misleading or sensationalized headlines that misrepresent the video's actual content. On the other hand, hate speech promotes negative attitudes and prejudices towards individuals or groups, leading to discrimination and division in society.

To lessen the aforementioned threats, YouTube has already prohibited uploading videos with misleading information and hate speech to its site. Although most of the study on YouTube is done to detect false content and hate speech in English and some other predominantly used languages, which has resulted in a rare amount of research on minority languages like Sinhala.

The prevailing trend of YouTube video ratings predominantly relying on metrics such as views and likes, often overlooking the video's actual content, has prompted the focus of this study. The primary objective is to introduce an alternative rating that centers on audio extraction from YouTube videos, enabling the identification of hate speech and inaccurate information in Sinhala YouTube videos. This innovative approach aims to inform users about the video's content integrity, empowering them to make more informed decisions on whether the video is worth watching..

II. LITERATURE REVIEW

*A. Hate Speech Detection*

Since a few years ago, hate speech identification has been a popular subject due to the rise in disinformation and violent content on social media. Every major social media network now faces a serious issue with hate [1] , [2].

A group of individuals can be attacked or denigrated based on their race, religion, national origin, gender, sexual orientation, age, disabilities, or illnesses, and violence against them can be encouraged by hate speech. The intricacies of language play a crucial role in the complex phenomena of hate speech, which is connected to interpersonal interactions within a community. Any communication that singles out or dehumanizes a certain group of individuals on the basis of traits such as race, ethnicity, gender, religion, sexual orientation, or handicap is referred to as such [3], [4].

In the [5] study, Sandaruwan et al. have stated the methodologies as lexicon-based and machine-learning-based. In their study, lexicon generation was the first step in the lexicon-based approach, and a corpus-based lexicon provided 76.3% accuracy for detecting hate, offensive, and neutral speech. Building a corpus of 3,000 evenly dispersed comments among hate, offensive, and neutral remarks was the first step in the machine learning technique. They were able to determine the optimum feature groups and models for Sinhala hate speech identification using this comment corpus. Their tests showed that the Multinomial Nave Bayes character trigram had the highest recall value, 0.84, and the best accuracy, 92.33%.

Samarasinghe et al. [3] in "Machine Learning Approach for the Detection of Hate Speech in Sinhala Unicode Text" have suggested a deep learning mechanism that utilizes two CNNs; in their study, a given corpus is first classified as hateful or not, and then that hateful corpus is classified according to the hate level. FastText word embedding is used to convert text data to numerical vectors. Since the data distribution within the set is uneven, the proposed model showed an accuracy of 83%, with an F1 score of 0.67.

Gamage et al. [4], in "Improving Sinhala Hate Speech Detection Using Deep Learning", have mentioned the importance of hate speech detection in native languages. They have investigated several embeddings and frequency-based features, BOW, n-gram, TF-IDF, etc., and as the final implementation, they have used deep learning and machine learning experiments. Out of the experiments with the ML algorithms LR, NB, RF, and SVM with frequency features, their results show that the LR classifier with Character n-gram (3, 5) achieved the highest 0.737 of F1 score. LR classifier with LaBSE achieved the highest F1 score of 0.642 when the word embeddings (word2vec, FastText) and contextual embeddings (XLMR, LaBSE)) were fed to the ML algorithms. Out of the deep learning, experiments carried out, CNN, RNN, GRU, LSTM, hybrid GRU-CNN, TNN, and transformer model (XLMR with LaBSE); XLMR model performed well with the F1 score of 0.764. Those well-performed models have obtained higher results than Samarasinghe et al. CNN model. The discussion says Sinhala hate speech detection should be directed towards the direction of a transformer-based approach by fine-tuning XLMR or other transformer models.

Currently, most of the techniques are applied to textual data to detect hate speech in the Sinhala language, In the context of "Detection of Hate Speech in Videos Using Machine Learning" [6] introduces a system that focuses on detecting hate speech on spoken content of the videos by Extracting audio from video, converting audio into text format and training machine learning models over text-based features and classify videos as normal or hateful. The study results indicate that the Random Forrest Classifier model provided the best results with an accuracy of 96%.

### B. False Information Detection

Misinformation refers to information that is disseminated widely without a basis in reality and without confirmation or clarification, whether intentionally or unintentionally [7].

The study 'Implementation of Speech to Text Conversion Using Hidden Markov Model', [8] was conducted to achieve audio-to-text transcription, with the objective of converting audio input from a user or computer into readable text. The researchers proposed the use of Hidden Markov Model (HMM) to address this challenge. The results showed that the HMM approach achieved a high level of accuracy, with 92.4% training accuracy, 93.6% testing accuracy, and 95.2% validation accuracy.

Researchers in a separate study [9] aimed to identify the similarity between any two texts. They utilized several similarity metrics, including Euclidean distance, Squared Euclidean distance, Manhattan distance, Chessboard distance, Bray Curtis Distance, and Canberra Distance. However, the researchers noted that "All the similarity measures gave almost the same results" and concluded that the best measurement largely depends on the data's characteristics.

Another study [10] highlights the limitations of traditional deep learning models for calculating text similarity, such as their inability to extract text semantics sufficiently, combine context, and understand polysemy. To overcome these drawbacks, the researchers propose the use of a Siamese network based on ALBERT for text similarity calculations.

Even with research on hate speech and false information in videos, the understanding of detecting hate speech and false information in YouTube videos is notably low. The research that have been done so far has mainly focused on detecting hate speech and false information in a few major languages, leaving out other languages that may not be widely used on the internet. This may cause many controversial situations as hate speech and false information in any minority language can lead to as bad an impact on some communities as they would in major languages, thus making it necessary to develop powerful hate speech detection and false information detection systems in any language. The existing gap in knowledge pertains to the lack of comprehensive research examining the effectiveness of current methodologies in detecting offensive speech, encompassing hate speech and false information, within YouTube videos. This deficiency is particularly pronounced when considering languages that are less prevalent on the internet, such as 'Sinhala', which constitutes a minority language.

## III. METHODOLOGY

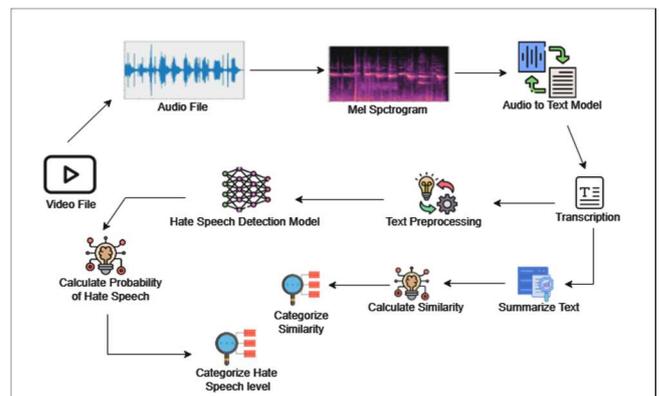

Fig. 1. System Diagram

*A. Datasets*

This research leverages a diverse array of datasets tailored to specific tasks. Notably, the OpenSLR/SLR52 dataset [11] is harnessed for the undertaking of automatic speech recognition, encompassing a substantial compilation of over 185,000 Sinhala audio samples, each meticulously paired with its corresponding transcription. In parallel, the Sinhala-Unicode-Hate-Speech Kaggle dataset [12] is adopted for the purpose of hate speech detection, featuring an extensive collection of more than 6000 sentences, accompanied by corresponding labels indicating the presence of hate speech. Furthermore, the Hamza-Ziyard/CNN-Daily-Mail-Sinhala dataset [13], accessible via the Hugging Face platform, is employed for sentence summarization endeavors. This dataset encompasses an assemblage of over 5000 Sinhala sentences, each accompanied by a concise summary.

*B. Audio extraction*

This study utilized YouTube API version 3 to gather crucial information regarding YouTube videos, encompassing metadata such as video title, duration, and prominent engagement metrics like view count and likes. Furthermore, the audio component was extracted from the videos employing the Pytube library. Subsequently, the extracted audio segments served as input data for the ASR (Automatic Speech Recognition) model employed in the study.

*C. Transcription*

Data preprocessing plays a crucial role in preparing the OpenSLR/SLR52 dataset for fine-tuning the Whisper model [14] in Sinhala language. To ensure seamless integration with the Whisper model, the audio files are initially converted to the WAV format, and their sample rate is standardized to 16kHz. To facilitate efficient training and prevent potential memory-related issues, the audio files are further segmented into 30-second chunks. Log-Mel spectrogram features are extracted from these segmented audio segments, serving as the representative audio input for the Whisper model. Meanwhile, the text transcriptions undergo meticulous cleaning, involving the removal of special characters, punctuation, and unnecessary whitespace. Furthermore, depending on the specifications of the particular Whisper variant, the text is tokenized into sub-word units, thereby creating an indispensable vocabulary essential for the training process.

Model fine-tuning constitutes a critical phase in the approach, wherein the preprocessed dataset is leveraged to adapt the Whisper model for the Sinhala text-to-speech task. The process commences by initializing the Whisper model with pre-trained weights, readily accessible from the Hugging Face model community or the OpenAI model repository. Subsequently, the dataset is partitioned into training and validation sets, enabling real-time monitoring of the model's performance throughout the training process. To assess the model's efficacy during training, the Word Error Rate (WER) is employed, gauging the disparity between the model's predicted output and the actual Sinhala text. Fine-tuning the model is executed using the Adam optimizer, with adjustments made to the learning rate to optimize convergence. An extensive exploration of diverse hyperparameters, including batch size, hidden layer size, number of attention heads, and dropout rate, is conducted to arrive at the most suitable configuration for both the dataset and the Whisper model.

*1) Hate Speech Detection:* During the initial phase, thorough preprocessing is applied to the Sinhala-Unicode Hate Speech dataset to ensure its appropriateness for refining the distilroberta-base model [15]. This process commences by meticulously collecting and structuring the dataset, comprising text samples categorized as either hate speech or non-hate speech. To enable effective model training and evaluation, the dataset is partitioned into three distinct subsets: a training set, essential for facilitating the model's learning journey; a validation set, crucial for fine-tuning parameters and monitoring progress; and a test set, pivotal for conducting a comprehensive assessment of the model's overall performance. Subsequently, a process known as tokenization is employed to transform the text. This involves breaking down the Sinhala text into sub-word tokens. The tokenizer's embedded vocabulary is then utilized to encode these tokenized representations into numerical input IDs. To ensure seamless integration with the model, identifiers denoting hate speech instances are meticulously converted into numeric values, with '1' signifying hate speech and '0' denoting non-hate speech instances. These transformed label representations are subsequently converted into tensors, aligning their architectural compatibility with the model's structure.

In the process of model training, a well-defined approach is followed. The foundational element is the integration of the distilroberta-base pre-trained model, which serves as the base for further refinement. A pivotal enhancement entails the incorporation of a specialized classification head, representing a key modification. This additional component is meticulously tailored for binary classification tasks, specifically aimed at effectively discerning between instances categorized as hate speech and those classified as non-hate speech. To ensure seamless adaptation, the output size of the classification head is adjusted to precisely correspond to the two distinct classes. It is important to underscore that, during the preliminary stages of training, the underlying model's weights are intentionally "frozen," preserving the valuable insights acquired during its pre-training phase.

Within the framework of training configuration, several critical factors are carefully determined. The selection of an appropriate batch size, the choice of a suitable optimizer, and the incorporation of a learning rate scheduler that dynamically fine-tunes the learning rate over time collectively contribute to a robust training regimen. Notably, the Binary Cross-Entropy loss function is judiciously employed to address the binary classification task's unique requirements.

The model's evolution progresses through a meticulous process termed fine-tuning, wherein it gradually acquires the capability to effectively differentiate between patterns associated with hate speech and those characteristics of non-hate speech within the training dataset. Rigorous monitoring of key validation metrics, such as loss and accuracy, acts as a safeguard against overfitting, ensuring the model's generalizability. Upon successful convergence, as determined through extensive testing on the validation set, the model's efficacy is subjected to a comprehensive evaluation of the test dataset. This evaluation encompasses a diverse set of performance metrics, including accuracy, precision, recall, and F1-score, which collectively provide a comprehensive and detailed understanding of the model's proficiency in accurately identifying Sinhala hate speech instances.

*2) Text Summarization:* The process commences with the Collection and Preprocessing of Datasets. In this initial phase, we obtain the CNN-Daily-Mail-Sinhala dataset, which comprises the Sinhala translation of English news articles alongside their corresponding summaries. To ensure the dataset's suitability for training, a meticulous preprocessing pipeline is established. Employing a tokenizer, the Sinhala text is segmented into sub-word units, aiding the model in breaking down the content into more manageable segments. Moreover, extraneous elements such as characters, punctuation, and special symbols are meticulously expunged, ensuring the data is purified and primed for training. Subsequently, the dataset is meticulously partitioned into distinct subsets for training, validation, and testing. This stratification not only exposes the model to a diverse range of data but also furnishes essential milestones to gauge its developmental trajectory.

The subsequent step involves the introduction of the model architecture. At the core of this methodology lies the BART-Large-XSUM pre-trained model [16], serving as the fundamental framework for Sinhala sentence summarization. Operating as a sequence-to-sequence model, BART encompasses a denoising autoencoder objective that aptly corresponds to the summarization task. The BART-Large-XSUM model, endowed with extensive pre-trained knowledge, furnishes a sturdy initial platform for the generation of summaries that exhibit both coherence and substantive significance.

In the subsequent phase, termed Fine-tuning, the BART-Large-XSUM model undergoes a targeted optimization procedure. This optimization is geared towards refining the model for Sinhala sentence summarization, leveraging its pre-trained weights. The model's configuration is adjusted to align with the Sinhala-specific sequence-to-sequence task, defining input and output parameters conducive to effective summarization. Employing the preprocessed dataset, this fine-tuning process takes shape. The quantification of dissimilarity between predicted and actual summaries is achieved through the application of a cross-entropy loss function.

## IV. RESULTS AND DISCUSSION

In this section, we present the results of our study, which centers on the implementation and evaluation of an audio extraction method, a transcription model for the Sinhala language, and their subsequent incorporation into a hate speech detection model. This section is divided into two parts: the performance of the transcription model and the efficiency of the integrated hate speech detection system.

### A. Transcription Model Performance

*1) Word Error Rate (WER):* The Word Error Rate (WER) is one of the most important metrics used to evaluate the precision of our transcription model. WER measures the disparity between the transcribed output and the true text. The transcription model obtained a WER of 48.99\%, according to our experiments. Fig. 2, below shows the WER by steps. While this number may appear excessive, it is important to consider the complexity of Sinhala phonetics and the variation in audio quality, both of which can contribute to transcription errors.

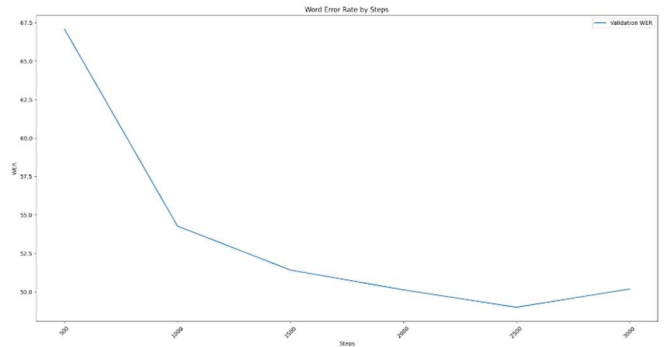

Fig. 2. Word Error Rate

*2) Training and Validation Losses:* To assess the training progress and generalization of the transcription model, the monitoring of training and validation losses was conducted throughout the training process. The training loss of the optimal model converged to a value of 0.125400, with the validation loss reaching 0.188352. These loss values collectively signify the model's adeptness in learning from the training data and its capability to reasonably generalize to validation samples which are not encountered during training. Fig. 3, below shows the training and the validation losses by Steps.

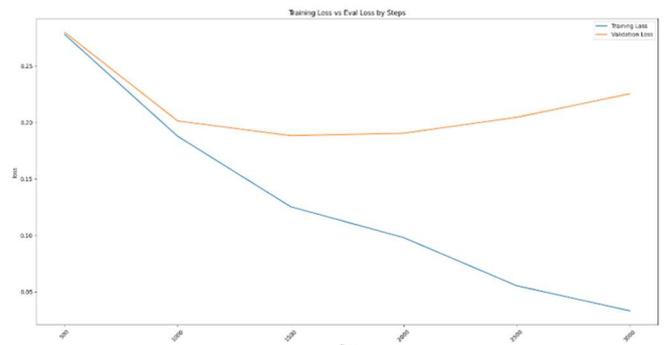

Fig. 3. Training and Validation Losses

The findings presented in Table I demonstrate the correlation between data set size and the performance metrics, specifically Word Error Rate (WER), Training Loss, and Validation Loss. Notably, as the data set size increases, there is a discernible enhancement in the performance of the model.

TABLE I.    TRANSCRIPTION MODEL PERFORMANCE

| Dataset Size | WER | Training Loss | Validation Loss |
|---|---|---|---|
| 3K | 54.19% | 0.1352 | 0.190916 |
| 6K | 51.67% | 0.1365 | 0.189311 |
| 12K | 48.99% | 0.1254 | 0.188352 |

### B. Performance of the Hate Speech Detection Model

Extensive efforts have been undertaken to ensure the equilibrium of training and testing datasets across all categories within the methodology. This balance allows for the firm consideration of **Accuracy** as a meaningful and significant metric for assessing the model's effectiveness.

The evaluation of the sentiment analysis model incorporates a diverse range of performance indicators, including **Precision**, **Recall**, and **F1-score**. These metrics hold particular importance due to the primary objective of identifying hate comments within a broader spectrum of comments, thus playing a critical role in the research.

Let's delve into the fundamental definitions of these metrics to gain further clarity:

- **True Positives (TP)**: Instances that have been correctly labeled as positive (indicating hate remarks).
- **True Negatives (TN)**: Instances that have been correctly labeled as negative (representing non-hateful comments).
- **False Positives (FP)**: Instances that have been incorrectly categorized as positive (or as hate comments).
- **False Negatives (FN)**: Instances that have been misclassified as negative (indicating instances where real hate remarks were missed).

Precision, Recall, and F1-score offer nuanced insights into the model's performance:

**Precision**: The proportion of correctly anticipated positive events out of all instances predicted as positive. A high precision value signifies the model's caution when classifying certain instances as hate speech.

**Recall**: The proportion of actual positive instances that were correctly predicted out of all positive instances. A high recall value indicates the model's effectiveness in capturing the majority of genuine hate comments.

The **F1-score**, which is the harmonic mean of Precision and Recall, provides a balanced measure that accounts for both false positives and false negatives. It offers a comprehensive assessment of the model's performance.

The outcomes of each experimental iteration are meticulously examined and presented in the extensive investigation, yielding an F1-score of 0.856, a precision value of 0.851, and a recall value of 0.861. Confusion Matrix for the distilroberta-base model is shown in the fig. 4.

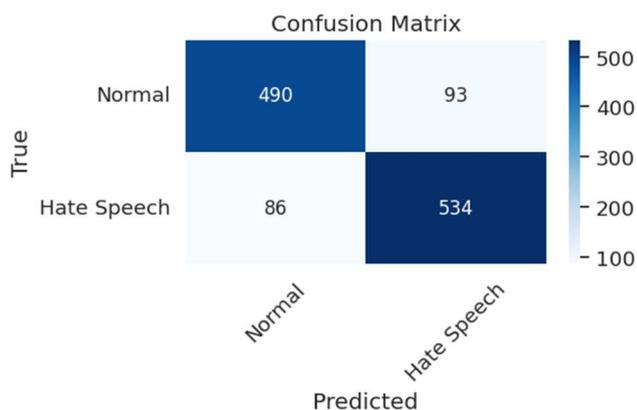

Fig. 4. Confusion Matrix for the distilroberta-base model

Based on the tests that were carried out and the results that were acquired, it has been determined that the RoBERTa model is the most optimal option. This conclusion arises from the observation that the LSTM model, as shown in Fig. 5 exhibited a notable vulnerability to overfitting.

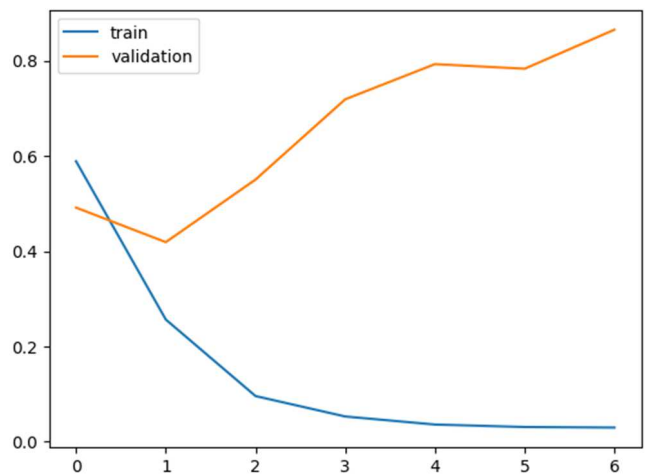

Fig. 5. Training and Validation Loss of the LSTM model

Table II shows the accuracy metrics for hate speech detection with the models LSTM and the distilroberta-base model.

TABLE II. HATE SPEECH MODEL PERFORMANCE

| Model | Accuracy | F1 score | Precision | Recall |
|---|---|---|---|---|
| distilroberta-base model | 0.851 | 0.856 | 0.851 | 0.861 |
| LSTM | 0.9426 | 0.9472 | 0.9651 | 0.9299 |

## V. CONCLUSION

In this study, the amazing power of pre-trained transformer models was demonstrated, and YouTube was compelled to act, to stop the spread of hate speech and false information because these problems have become major global challenges. There is still a study gap despite the efforts to restrict such content, particularly when it comes to non-English content like Sinhala. This study presents a novel approach by developing a comprehensive rating system for YouTube videos and focusing on the incidence of false information and hate speech in Sinhala audio. The approach evaluates titles, descriptions, and audio content to assign ratings based on the similarity of false information and the presence of offensive language. The method involves extracting the audio from the video using the Pytube library, audio transcription done using the Whisper model that has been fine-tuned, and then the transcript is used to identify hate speech using the distilroberta-base model that has been tuned as a text classification model which performed well with an F1-score of 0.856, precision of 0.851 and a recall value of 0.861. Based on the conducted tests and results, the RoBERTa model emerges as the optimal choice, primarily due to the significant overfitting vulnerability observed in the LSTM model. In order to obtain a similarity score with the title, text summarization is done to the transcript using the BART-Large-XSUM model that has been finetuned. According to the Whisper model, the WER is 48.99%.

While the suggested solution primarily focuses on enhancing the functionality of the YouTube platform, it is important to highlight that its versatility extends beyond this specific application. The innovations and features introduced in this solution can be seamlessly integrated into various other video streaming platforms, making it a valuable and adaptable solution for a wider range of digital media platforms.

As for future work, using hyper-parameter optimization and fine-tuning various transformer models, the rating system's accuracy could be improved by working with linguists to better recognize subtleties in hate speech and false information in Sinhala.